\documentclass[aps,pra,twocolumn,floatfix,showpacs]{revtex4-2}
\usepackage{amsmath,amssymb,amsfonts}
\usepackage{graphicx}
\usepackage{bm}

\begin{document}

\title{Erasure of entanglement-locked energy in multi-cell quantum batteries
       through decoherence}

\author{Haniyeh Hajizade}
\author{Vahid Ameri}
\email[]{vahameri@gmail.com}
\author{Alidad Askari}
\affiliation{Department of Physics, Faculty of Science, University of Hormozgan, Bandar Abbas, Iran}

\date{\today}

\begin{abstract}
Charging a quantum battery through a pumped charger can lock a substantial part of its
stored energy inside the bipartite correlations that grow between the constituent cells,
and this \emph{correlation ergotropy}, being the energy that no local single-cell
operation can reach and that only collective operations can recover, is the object of
the present study. Following the charging trajectory of two coupled three-level cells,
we find that the correlation ergotropy closely tracks the logarithmic negativity, which
shows that the locked energy is carried by genuine bipartite entanglement rather than by
coherent collective dynamics alone, and that an equal-strength coherent non-entangling
drive produces no such locking at all. For a pure two-cell battery we establish an exact,
basis-independent closed form in which the locked energy depends only on the sorted
eigenvalues of the reduced cell and the cell energy levels, revealing that the locked
fraction rises from zero at the separable corners of the Schmidt simplex to unity at the
symmetric maximally entangled state and remains invariant under local unitaries.
Decoherence and dephasing, which destroy the correlations, do \emph{not} release this
energy but erase it, so extraction must be collective and must take place while the
correlations persist, while cold damping leaves the effect intact and thermal occupation
suppresses it. These results give a resource-theoretic reading of correlations in
quantum batteries, sharpen the line between genuine entanglement and coherent collective
dynamics, and point to experimentally accessible conditions for correlated-charge and
collective-extraction protocols.
\end{abstract}


\maketitle

\section{Introduction}\label{sec:intro}

Quantum batteries are devices that store and release energy using quantum resources
such as coherence and correlations \cite{Alicki2013,Binder2015,Binder2015NJP,Campaioli2018,Campaioli2024}. The
paradigm was introduced by Alicki and Fannes \cite{Alicki2013}, who showed that
entangled work can increase the extractable work from a composite system. Since then,
a large body of work has examined how coherent driving, collective coupling, and
genuine quantum correlations affect the stored energy and the charging power
\cite{Binder2015,Campaioli2018,Perarnau2015,Ferraro2018,Andolina2019PRL,Andolina2019,
Shi2022,Delmonte2019,Gyhm2022,Quach2022,Gherardini2020,Chen2020,Purkait2026}.

A central figure of merit is the \emph{ergotropy} \cite{Allahverdyan2004}: the
maximum work that can be extracted from a state by unitary operations. For a battery
composed of many cells, extractability hinges on whether the operations
act on individual cells (local operations) or on the whole battery (collective
operations). Correlations between the cells store energy that is inaccessible to
local unitaries, giving rise to \emph{correlation ergotropy}, also called quantum
ergotropy \cite{Francica2019,Francica2020,Shi2022,dElRey2026,Wang2026}. The connection between
entanglement and extractable work has been studied in several settings
\cite{Francica2020,Shi2022,Mitra2021,Yang2024}, and it is now understood that quantum
coherence in the battery or entanglement with a charger is a necessary resource for
nonzero extractable work during charging \cite{Shi2022}.

A question that keeps recurring is whether correlations can be used to
\emph{overcharge} a battery: if the charger entangles the cells, can one store more
energy and, once the correlations fade, extract more than one otherwise could? The two
parts of this question have different answers, and we address both. Correlations do
allow the storage of energy that no local operation can reach, but that energy is
\emph{locked}: it is available only to collective operations, and when decoherence
destroys the correlations it is erased rather than released. Correlations never raise
the total stored energy: entanglement reconfigures where the energy is accessible but
does not create more of it. These statements follow directly from the definition of
ergotropy together with direct numerical integration of the master equation. What we add is a concrete, experimentally relevant model in which
they hold, and a clean identification of the resource (entanglement) behind the locked
energy.

The setting is deliberately minimal: a battery of two (and later $N$) three-level atoms
(qutrits) charged by a single pumped three-level charger. It is rich enough to support
genuine bipartite entanglement between the cells, yet small enough for exact numerical
treatment with QuTiP \cite{Qutip}. Along the charging trajectory we compute the global
and local ergotropies, the logarithmic negativity, the mutual information, and the
stored energy, and compare the entangling charger with a coherent, non-entangling
drive. The last step is to examine how the effect responds to dissipation and finite
temperature, and how it scales with the number of cells.

The paper is organized as follows. In Sec.~\ref{sec:model} we introduce the model and
the figures of merit. In Sec.~\ref{sec:results} we present the main results: the
entanglement origin of correlation ergotropy (Sec.~\ref{sec:origin}), an exact
closed form for the correlation ergotropy of pure two-cell states
(Sec.~\ref{sec:analytical}), the contrast with a coherent non-entangling drive
(Sec.~\ref{sec:coherent}), the erasure by decoherence (Sec.~\ref{sec:decoherence}), collective extraction
(Sec.~\ref{sec:extraction}), scaling (Sec.~\ref{sec:scaling}), and dissipation
robustness (Sec.~\ref{sec:dissip}). We close with a discussion in
Sec.~\ref{sec:discussion}.

\section{Model and figures of merit}\label{sec:model}

\subsection{Charging Hamiltonian}\label{sec:hamiltonian}

Each cell is a three-level system (qutrit) with states
$\{|0\rangle,|1\rangle,|2\rangle\}$ and Hamiltonian
\begin{equation}
H_{\rm cell} = \omega\,\bigl(|1\rangle\langle1|+2|2\rangle\langle2|\bigr),
\label{eq:hcell}
\end{equation}
where $\omega$ sets the energy scale. We take $\omega=1$ throughout. The transition
operators are $\sigma^{01}=|0\rangle\langle1|$, $\sigma^{12}=|1\rangle\langle2|$, and
$\sigma^{02}=|0\rangle\langle2|$.

For two cells $A,B$ and a charger $C$ (all qutrits), the total Hamiltonian during
charging is
\begin{equation}
\begin{split}
H_{\rm ch}&=H_A+H_B+H_C
+\Omega\bigl(\sigma_C^{02}+\sigma_C^{20}\bigr)\\
&\quad+\lambda\bigl(\sigma_A^{01}+\sigma_A^{10}+\sigma_B^{01}+\sigma_B^{10}\bigr)
\bigl(\sigma_C^{01}+\sigma_C^{10}\bigr),
\end{split}
\label{eq:hcharge}
\end{equation}
where $\Omega$ is the pump amplitude driving the $|0\rangle\leftrightarrow|2\rangle$
transition of the charger and $\lambda$ is the charger--cell coupling. The pump
injects energy into the charger, which transfers it to both cells, generating
correlations between them. Unless stated otherwise we use
$\Omega=0.7\omega$ and $\lambda=0.4\omega$.

For the coherent, non-entangling drive we pump each cell directly and in phase,
\begin{equation}
H_{\rm coh}=H_A+H_B+\Omega\sum_{i=A,B}\bigl(\sigma_i^{02}+\sigma_i^{20}\bigr),
\label{eq:hcoherent}
\end{equation}
which leaves the two cells in a product state at all times. This protocol has the
same pump amplitude and injects energy at a comparable rate, but it does not create
entanglement between the cells.

\subsection{Figures of merit}\label{sec:figures}

The ergotropy of a state $\rho$ with respect to a Hamiltonian $H$ is
\cite{Allahverdyan2004}
\begin{equation}
\mathcal{W}(\rho,H)=\mathrm{Tr}\!\left[H\rho\right]
-\mathrm{Tr}\!\left[H\rho^{\rm p}\right],
\label{eq:ergo}
\end{equation}
where $\rho^{\rm p}$ is the \emph{passive} state, obtained by arranging the
populations of $\rho$ in decreasing order of the corresponding eigenvalues of $H$.
For the two-cell battery we define the \emph{global} ergotropy
\begin{equation}
\mathcal{W}_{\rm g}=\mathcal{W}(\rho_{AB},H_A+H_B),
\label{eq:wglobal}
\end{equation}
the maximum work extractable by collective (global) operations, and the \emph{local}
ergotropy
\begin{equation}
\mathcal{W}_{\rm l}=\mathcal{W}(\rho_A,H_A)+\mathcal{W}(\rho_B,H_B),
\label{eq:wlocal}
\end{equation}
the maximum work extractable by acting on each cell independently (also called
parallel ergotropy \cite{Castellano2025}). The difference
\begin{equation}
\mathcal{W}_{\rm corr}\equiv\mathcal{W}_{\rm g}-\mathcal{W}_{\rm l}
\label{eq:wcorr}
\end{equation}
is the correlation ergotropy \cite{Francica2019,Shi2022,Simon2025}, the energy stored in
correlations that local operations cannot reach. This quantity is the \emph{ergotropic
gap} between the global and local ergotropies, a nonnegative witness of energy-relevant
correlations: because local operations form a subset of global ones,
$\mathcal{W}_{\rm g}\ge\mathcal{W}_{\rm l}$, with equality for a product state.

We quantify the entanglement between the two cells by the logarithmic negativity
\begin{equation}
\mathcal{N}(\rho_{AB})=\log\bigl\|\rho_{AB}^{T_A}\bigr\|_1,
\label{eq:neg}
\end{equation}
where $T_A$ denotes partial transposition with respect to subsystem $A$. For a pure
two-cell state with Schmidt coefficients $\{\lambda_n\}$, this reduces to
$\mathcal{N}=2\log(\sum_n\lambda_n^{1/2})$. The mutual information is
$I(\rho_{AB})=S(\rho_A)+S(\rho_B)-S(\rho_{AB})$, with $S$ the von Neumann entropy.

\subsection{Open-system dynamics}\label{sec:open}

For the dissipation studies we couple each cell to an independent thermal bath
described by the Lindblad master equation. For a cell with damping rate $\gamma$ and
thermal occupation $n_{\rm th}$, the collapse operators are
$\sqrt{\gamma(n_{\rm th}+1)}\,\sigma^{01}$, $\sqrt{\gamma(n_{\rm th}+1)}\,\sigma^{12}$
and the corresponding raising operators. At $n_{\rm th}=0$ only the lowering operators
contribute, while the raising operators appear only at finite temperature. The dynamics
are integrated exactly with the QuTiP library \cite{Qutip} using dense evolution. The
total Hilbert-space dimension is $3^{N+1}$ for $N$ cells plus one charger, so for
$N\le4$ the full state is propagated exactly. Convergence was checked by increasing the
number of time points until the stored energy and the ergotropies changed by less than
$10^{-6}$ in relative terms.

\section{Results}\label{sec:results}

The results below build the picture announced in the Introduction: collective charging
entangles the cells and locks part of the stored energy in correlations, the locked
energy is carried by entanglement and must be extracted collectively, and decoherence
erases rather than releases it.

\subsection{Correlation ergotropy is carried by entanglement}\label{sec:origin}

\begin{figure}[t]
\includegraphics[width=\columnwidth]{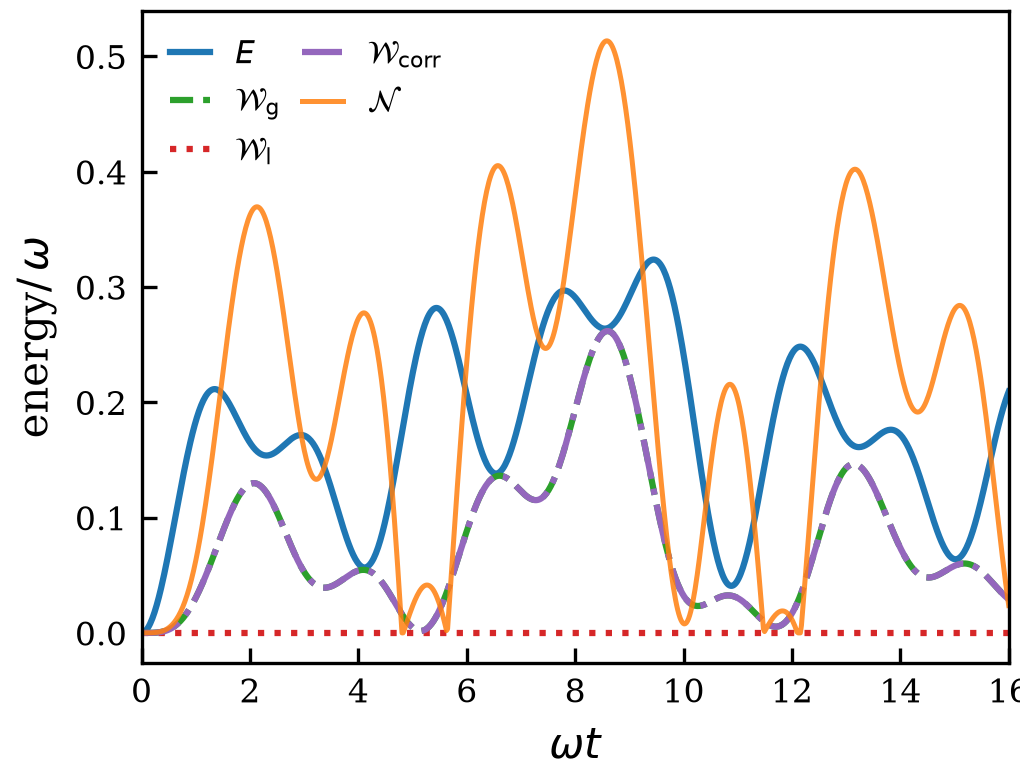}
\caption{Charging trajectory of two qutrit cells with a quantum (entangling) charger
($\Omega=0.7\omega$, $\lambda=0.4\omega$): stored energy $E$, global ergotropy
$\mathcal{W}_{\rm g}$, local ergotropy $\mathcal{W}_{\rm l}$, correlation ergotropy
$\mathcal{W}_{\rm corr}=\mathcal{W}_{\rm g}-\mathcal{W}_{\rm l}$, and logarithmic
negativity $\mathcal{N}$. At the correlation peak $\mathcal{W}_{\rm l}=0$ while
$\mathcal{W}_{\rm g}>0$, and $\mathcal{W}_{\rm corr}$ tracks $\mathcal{N}$.}
\label{fig:traj}
\end{figure}

Figure~\ref{fig:traj} shows the stored energy $E$, the global and local ergotropies,
the correlation ergotropy, and the logarithmic negativity along the charging
trajectory of the two-cell model with the quantum (entangling) charger. Two
features stand out:

\begin{itemize}
\item At the correlation peak, $\mathcal{W}_{\rm l}=0$ while $\mathcal{W}_{\rm g}>0$:
the stored energy is \emph{entirely} locked in correlations and inaccessible to local
operations.
\item The correlation ergotropy closely tracks the logarithmic negativity. Linear
regression gives Pearson $r=0.890$ ($p\simeq2\times10^{-270}$) and Spearman $0.912$,
establishing that, along this charging trajectory, the locked energy is strongly
correlated with the two-cell entanglement.
\end{itemize}

\begin{figure}[t]
\includegraphics[width=\columnwidth]{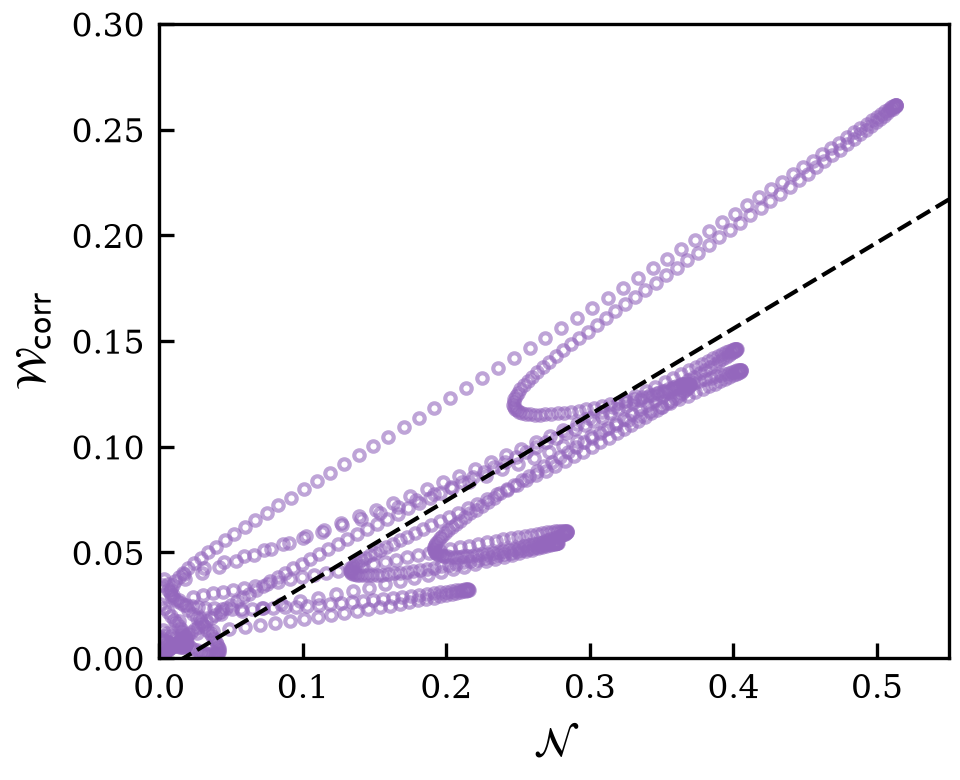}
\caption{Correlation ergotropy $\mathcal{W}_{\rm corr}$ versus logarithmic negativity
$\mathcal{N}$ along the charging trajectory of two qutrit cells. The near-linear
relation (dashed line, Pearson $r_{\rm P}=0.89$, Spearman $r_{\rm S}=0.91$) confirms
that the locked energy is carried by genuine bipartite entanglement.}
\label{fig:entang}
\end{figure}

These observations are summarized in Fig.~\ref{fig:entang}, which shows the
near-linear relation between $\mathcal{W}_{\rm corr}$ and $\mathcal{N}$ along the
trajectory. They show that the resource generating the correlation ergotropy in
this protocol is the bipartite entanglement of the cells, not merely coherent
collective dynamics.

It is worth noting that this is a dynamical, trajectory-specific statement and not
a general implication of $\mathcal{W}_{\rm corr}>0$. Correlation ergotropy is \emph{not}
an entanglement witness: separable, classically correlated states can carry nonzero
$\mathcal{W}_{\rm corr}$. For example, the separable mixture
$\rho=\tfrac12(|00\rangle\langle00|+|11\rangle\langle11|)$ has zero logarithmic
negativity. Each reduced cell is already passive, so $\mathcal{W}_{\rm l}=0$, while a
global operation maps the $|11\rangle$ component down to $|01\rangle$ at no local
cost, giving $\mathcal{W}_{\rm g}=\omega/2$ and hence $\mathcal{W}_{\rm corr}=\omega/2>0$.
The contrast that identifies entanglement as the operative resource here is therefore
not the mere sign of $\mathcal{W}_{\rm corr}$ but its \emph{quantitative} tracking of
$\mathcal{N}$ under the entangling charger together with its vanishing under the
coherent drive.

\subsection{Analytical correlation ergotropy for pure two-cell states}\label{sec:analytical}

The correlation ergotropy of a \emph{pure} two-cell battery admits a simple exact
form, which makes its dependence on the entanglement structure explicit and grounds the
numerical results above. Consider a pure state of two identical $d$-level cells with
ascending single-cell energy levels $e_0\le e_1\le\cdots\le e_{d-1}$ and Schmidt
decomposition
\begin{equation}
|\psi\rangle=\sum_{n=0}^{d-1}\sqrt{\lambda_n}\,|n_A\rangle|n_B\rangle,
\label{eq:schmidt}
\end{equation}
with $\sum_n\lambda_n=1$. Let $p_0\ge p_1\ge\cdots\ge p_{d-1}$ denote the Schmidt
coefficients $\lambda_n$ sorted in decreasing order. No assumption is made about the
orientation of the Schmidt bases $\{|n_A\rangle\},\{|n_B\rangle\}$ relative to the
energy basis.

Three exact statements follow from the definition of ergotropy. First, because a
global unitary can map any pure state onto the two-cell ground state, the global
ergotropy equals the stored energy,
\begin{equation}
\mathcal{W}_{\rm g}=E.
\label{eq:analytical_E}
\end{equation}
Second, the two reduced cells share the same spectrum $\{\lambda_n\}$ and therefore the
same passive energy $\sum_k p_k e_k$, so the local ergotropy, the sum of the two
single-cell ergotropies, is
\begin{equation}
\mathcal{W}_{\rm l}=E-2\sum_k p_k e_k.
\label{eq:analytical_Wl}
\end{equation}
Subtracting, the correlation ergotropy takes the compact, basis-independent closed form
\begin{equation}
\boxed{\;\mathcal{W}_{\rm corr}=\mathcal{W}_{\rm g}-\mathcal{W}_{\rm l}
=2\sum_k p_k e_k\;}
\label{eq:analytical_wcorr}
\end{equation}
with $p_0\ge p_1\ge\cdots$ and $e_0\le e_1\le\cdots$. The stored energy $E$ cancels
exactly, so the locked energy depends only on the sorted Schmidt coefficients and the
energy levels, and Eq.~\eqref{eq:analytical_wcorr} is exact for \emph{every} pure
two-cell state, independently of the Schmidt basis.

\begin{figure}[t]
\includegraphics[width=\columnwidth]{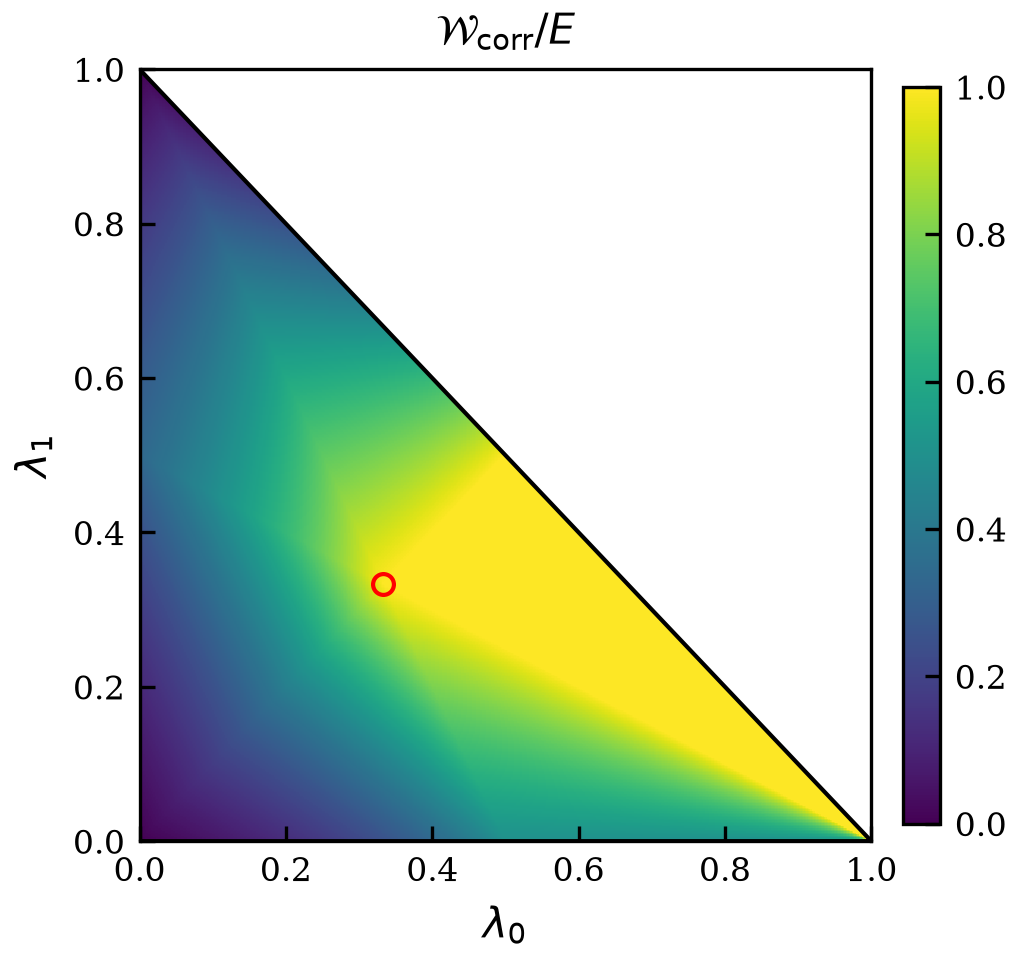}
\caption{Ratio $\mathcal{W}_{\rm corr}/E$ of a pure two-cell qutrit battery over the
Schmidt simplex $(\lambda_0,\lambda_1,\lambda_2)$ [with
$\lambda_2=1-\lambda_0-\lambda_1$], from the exact result
Eq.~\eqref{eq:analytical_wcorr}. The locked fraction vanishes at the separable
corners and reaches unity at the symmetric maximally entangled state
$(|00\rangle+|11\rangle+|22\rangle)/\sqrt3$ (red circle).}
\label{fig:schmidt}
\end{figure}

For the equally spaced levels of Eq.~\eqref{eq:hcell}, $e_n=n\omega$, the closed form
reduces to $\mathcal{W}_{\rm corr}=2\omega(p_1+2p_2)$ for $d=3$. It vanishes for
product states and reaches $\mathcal{W}_{\rm corr}=E=2\omega$ for the symmetric
maximally entangled state $(|00\rangle+|11\rangle+|22\rangle)/\sqrt3$. The ratio
$\mathcal{W}_{\rm corr}/E$ is mapped over the full Schmidt simplex in
Fig.~\ref{fig:schmidt}, which shows that the locked fraction grows from $0$ at the
separable corners to $1$ at the symmetric center. We verified
Eqs.~\eqref{eq:analytical_E}--\eqref{eq:analytical_wcorr} against exact numerical
ergotropy over $2000$ random Schmidt states to machine precision (maximum error
$<10^{-14}$), and for arbitrary $d$ over random generic pure states (maximum error
$<10^{-14}$). This analytical characterization shows that the correlation ergotropy
generated by the charging dynamics of Sec.~\ref{sec:origin} is bounded by the closed
form of Eq.~\eqref{eq:analytical_wcorr}, and approaches it as the two-cell state
becomes pure and symmetric. The same argument extends to a symmetric $N$-cell battery
in the GHZ-type state $|\psi\rangle=\sum_n\sqrt{\lambda_n}|n\cdots n\rangle$, whose
reduced cells all share the spectrum $\{\lambda_n\}$, giving
$\mathcal{W}_{\rm corr}=N\sum_k p_k e_k$ exactly.

\subsection{A coherent non-entangling drive produces no locked energy}\label{sec:coherent}

The quantum (entangling) charger stands in clear contrast to the coherent drive of
Eq.~\eqref{eq:hcoherent}, which keeps the cells in a product state
($\mathcal{N}=0$) at all times. For the coherent drive the local and global
ergotropies coincide, $\mathcal{W}_{\rm l}=\mathcal{W}_{\rm g}$, and hence
$\mathcal{W}_{\rm corr}=0$ for all time: every unit of stored energy is locally
accessible.

\begin{figure}[t]
\includegraphics[width=\columnwidth]{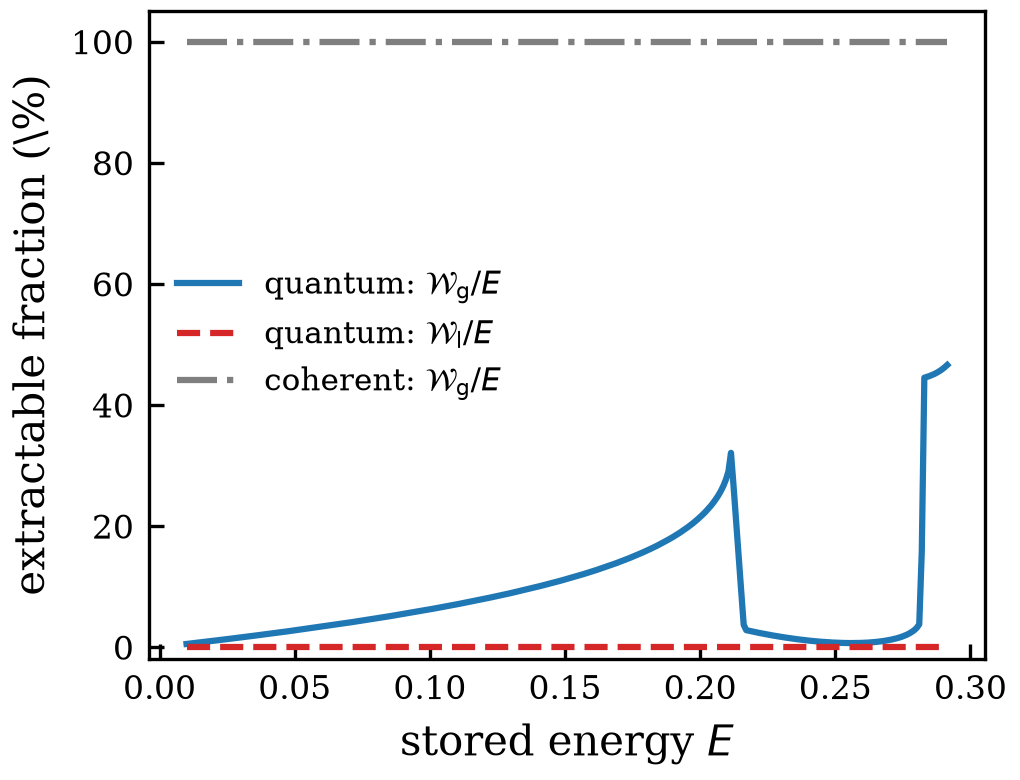}
\caption{Extractable fraction at matched stored energy $E$ for the quantum
(entangling) charger and the coherent (non-entangling) drive. The coherent drive
extracts all of its stored energy locally ($\mathcal{W}_{\rm g}=\mathcal{W}_{\rm l}=E$),
while the entangled charger leaves $\mathcal{W}_{\rm l}=0$ and reaches at most
$\mathcal{W}_{\rm g}/E\simeq47\%$ over the range shown. Entanglement transforms locally accessible energy
into correlation ergotropy rather than increasing it.}
\label{fig:fair}
\end{figure}

To compare the two protocols fairly, we fix the \emph{stored} energy to a common
value $E$ and compare the extractable fractions $\mathcal{W}_{\rm g}/E$ and
$\mathcal{W}_{\rm l}/E$ (Fig.~\ref{fig:fair}). At any fixed $E$, the coherent drive
extracts all of $E$ locally, whereas the entangled charger leaves
$\mathcal{W}_{\rm l}=0$ and, over the matched-energy range shown in
Fig.~\ref{fig:fair}, reaches at most $\mathcal{W}_{\rm g}/E\simeq47\%$ along the charging
branch (at the tuned correlation peak of Sec.~\ref{sec:extraction} the collective fraction
is instead $\simeq99\%$). The
entangled charger therefore does \emph{not} increase the extractable energy. Its
effect is a resource transformation: it converts locally accessible energy into
correlation ergotropy that is locked for local operations and accessible only
collectively. In resource-theoretic terms the two drives store energy in different
resources. The coherent drive deposits energetic coherence within each cell, a
single-cell resource that local unitaries convert directly into work, so its stored
energy is locally extractable. The entangling charger instead stores the same energy in
bipartite correlations, which are invisible to local operations. This matches recent
resource decompositions of ergotropy, in which locally extractable work is carried by
energetic coherence \cite{dElRey2026} and coherence acts as the fuel for extractable
work \cite{Jiang2026,Tarif2026,Oularabi2025}.

\subsection{Decoherence erases, it does not release}\label{sec:decoherence}

To test this, we took an entangled state with substantial correlation ergotropy and
let it undergo pure local dephasing, which destroys the correlations while conserving
the local populations and hence the stored energy. The local ergotropy stayed at
$\mathcal{W}_{\rm l}=0$ throughout, while the correlation ergotropy decayed
monotonically (e.g.\ $0.262\to0.130$ over the dephasing time). Losing the
correlation, in other words, does \emph{not} turn locked energy into locally
extractable work. It simply erases it. This is the quantitative refutation of the
``overcharge-then-unplug-to-release'' hypothesis. The effect is also distinct from the
noise-assisted coherent work reported for states with energetic coherence
\cite{Arjmandi2026}: here local dephasing preserves the populations, so the coherent
part of $\mathcal{W}_{\rm l}$ remains zero and only the bipartite correlation energy
$\mathcal{W}_{\rm corr}$ is lost.

\subsection{Collective extraction unlocks the correlation energy}\label{sec:extraction}

The locked energy is extracted by a \emph{collective} (global) operation performed
while the correlation is present. The maximum such work is by definition the global
ergotropy $\mathcal{W}_{\rm g}$. At the tuned operating point ($\Omega=0.70$,
$\lambda=0.40$) we find $E=0.264$, $\mathcal{W}_{\rm g}=0.262$ ($99\%$ of stored
energy), and $\mathcal{W}_{\rm l}=0$. The maximum is achieved by a single collective
operation: the global passivation unitary maps the two-cell state onto its passive
state with respect to $H_A+H_B$ and releases $\mathcal{W}_{\rm g}=0.262$ directly.
Importantly, the collective operation must be performed \emph{before}
the correlations decay. Waiting for decoherence to remove the correlation erases the
locked energy instead. Since the operating-point state is already nearly
globally passive ($\mathcal{W}_{\rm g}/E\simeq0.99$), this single collective
passivation recovers almost all of the stored energy.

We also checked whether an additional free collective step could first turn this
energy into a locally active form that could then be discharged locally. An
energy-conserving gate, a unitary that commutes with $H_A+H_B$ (a block-diagonal
rotation within degenerate energy manifolds), releases no work by construction, and
numerically it cannot raise the local ergotropy of the operating-point state above
zero, so no energy-conserving ``unlock'' exists. Conversely, unrestricted global
unitaries that inflate the local ergotropy beyond $\mathcal{W}_{\rm g}$ do so only by
injecting energy and are unphysical as free operations. The collective advantage is
therefore correctly benchmarked by $\mathcal{W}_{\rm g}$, which already captures
almost all of the stored energy in this model.

\subsection{Scaling with the number of cells}\label{sec:scaling}

\begin{figure}[t]
\includegraphics[width=\columnwidth]{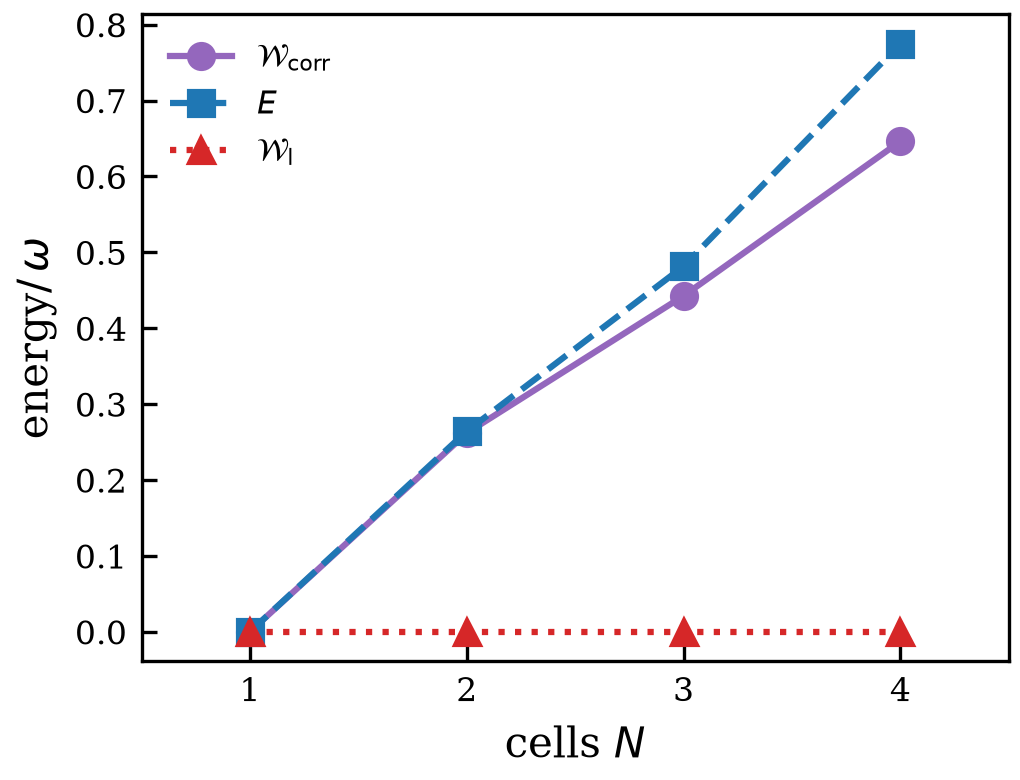}
\caption{Correlation ergotropy $\mathcal{W}_{\rm corr}$ (with the charger traced out),
stored energy $E$, and local ergotropy $\mathcal{W}_{\rm l}$ versus the number $N$ of
qutrit cells charged by a single pumped charger ($\Omega=0.7\omega$, $\lambda=0.4\omega$).
The correlation ergotropy grows $\sim N^{1.3}$ and the stored energy $\sim N^{1.5}$,
while the local ergotropy remains zero.}
\label{fig:scaling}
\end{figure}

When $N$ qutrit cells are charged by a single pumped charger and the charger is traced
out, the correlation ergotropy at its peak grows roughly as $N^{1.3}$ and the stored
energy as $N^{1.5}$ (Fig.~\ref{fig:scaling}), while the local ergotropy remains zero.
The collective fraction $\mathcal{W}_{\rm g}/E$ is close to unity for $N=2$ but settles
near $0.85$ for larger $N$, reflecting a residual entanglement between the charger
and the cells at the peak. Two caveats apply. First, these exponents are read off a
small window at a fixed charge time, and because the charging is oscillatory and the
peak drifts with $N$, they indicate a trend rather than an asymptotic law. Second, the
correlation ergotropy is only mildly superlinear, and its tilt is modest,
and the essential message is that collective charging produces energy whose extraction
requires collective operations, in line with the collective-charging advantage studied
in the literature
\cite{Perarnau2015,Ferraro2018,Campaioli2018PRX,Andolina2019PRL,Andolina2019,Campaioli2018}
and with the ergotropy--entanglement relation of Ref.~\cite{Zhi2025}.

\subsection{Dissipation and finite-temperature robustness}\label{sec:dissip}

\begin{figure}[htbp]
\includegraphics[width=0.86\columnwidth]{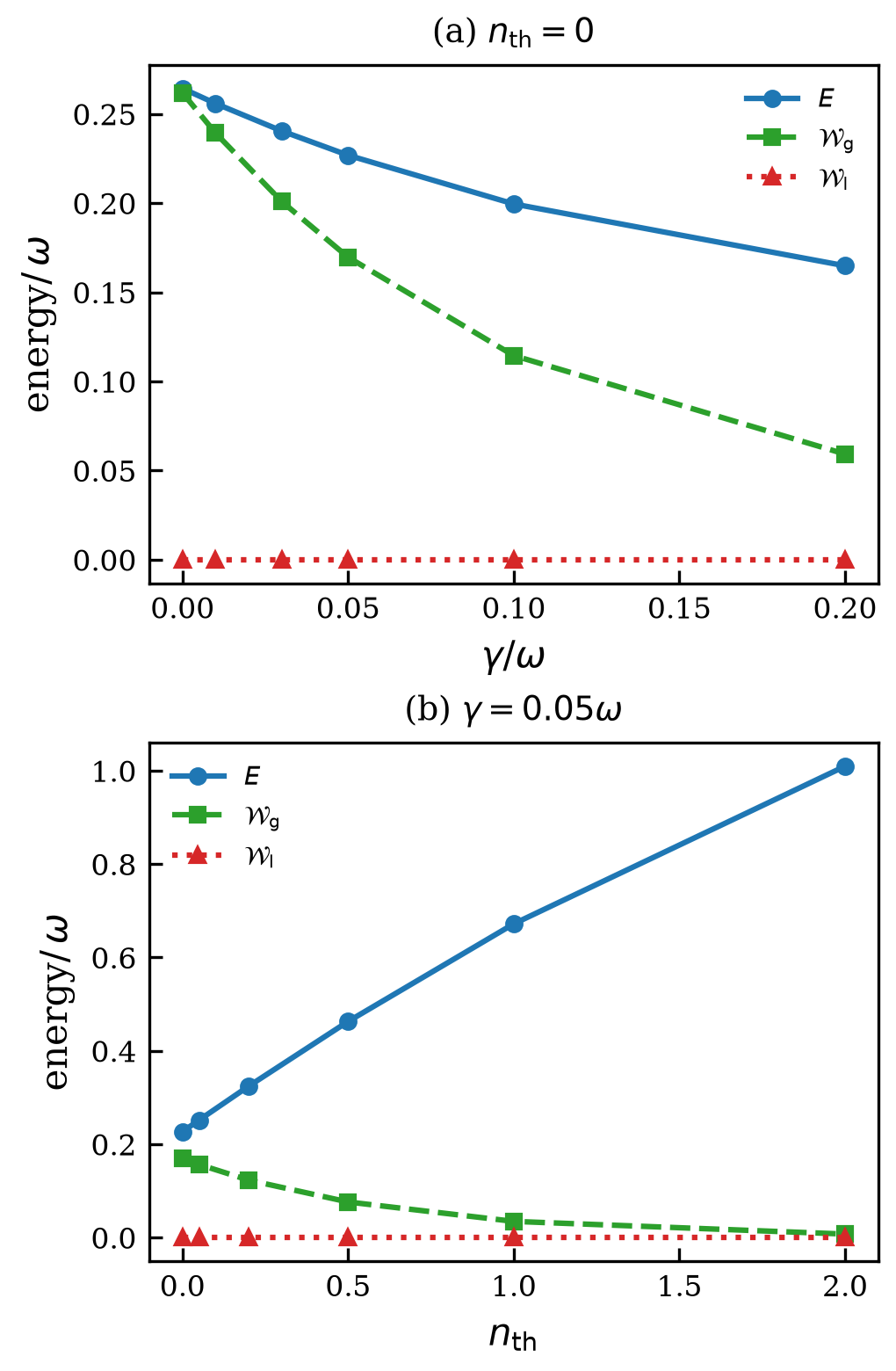}
\caption{Dissipation and finite-temperature robustness of the stored energy $E$,
global ergotropy $\mathcal{W}_{\rm g}$, and local ergotropy $\mathcal{W}_{\rm l}$.
(a) Cold bath ($n_{\rm th}=0$): $\mathcal{W}_{\rm g}$ survives moderate damping
$\gamma$, while $\mathcal{W}_{\rm l}$ remains zero. (b) At $\gamma=0.05\omega$, raising
the thermal occupation $n_{\rm th}$ rapidly suppresses $\mathcal{W}_{\rm g}$.}
\label{fig:dissip}
\end{figure}

When a thermal bath is included during charging (damping $\gamma$, thermal occupation
$n_{\rm th}$), the collective extractability $\mathcal{W}_{\rm g}/E$ is robust to
cold damping: $\gamma=0\to99\%$, $\gamma=0.03\to83\%$, $\gamma=0.05\to75\%$ at
$n_{\rm th}=0$ (Fig.~\ref{fig:dissip}a). It is, however, strongly suppressed by
temperature: at $\gamma=0.05\omega$, increasing $n_{\rm th}$ from $0$ to $2$ reduces
$\mathcal{W}_{\rm g}/E$ from $75\%$ to $0.7\%$ (Fig.~\ref{fig:dissip}b). The effect
therefore survives in cold (cryogenic or ion-trap) environments but is lost at high
temperature. This provides a falsifiable experimental prediction, and is consistent
with the observation that open-system charging and work extraction are sensitive to
the reservoir, including its spatial correlations \cite{Barra2019,Ahuja2025,Kamin2020}.

\section{Discussion and conclusions}\label{sec:discussion}

Taken together, our results paint a resource-theoretic picture of correlations in
quantum batteries. Entanglement built up during collective charging turns stored energy
into correlation ergotropy: energy that is locked away from local operations and can be
recovered only collectively. This locked energy tracks the entanglement along the
charging trajectory, grows with system size, and survives cold
environments while disappearing at high temperature. A coherent, non-entangling drive,
by contrast, stores energy that is fully accessible but locks none of it. Perhaps most
strikingly, when decoherence destroys the correlations, the locked energy is erased
rather than released. A naive ``overcharge-then-release'' strategy therefore fails, and
extraction must be performed collectively while the correlations are still present.

This helps settle a question that has been debated in the field: whether the advantage
of collective charging comes from genuine quantum correlations or merely from coherent
collective dynamics \cite{arXiv2601,Gyhm2022,Cavaliere2026,Jiang2026}. In our model the
\emph{locked} energy is driven by entanglement, even though entanglement does not
increase the stored energy itself. Our finite-dimensional result complements a recent
bosonic treatment in which the passive (locked) part of the stored energy is likewise
set by charger--battery entanglement \cite{dElRey2026}, confirming that the global
versus local distinction we exploit is a general feature of correlated batteries rather
than an artifact of qutrits. We also identify the regime (a charger--cell coupling
comparable to the pump, and low temperature) in which the correlated-charge /
collective-extract strategy is most advantageous, together with an experimentally
accessible protocol.

\subsection{Resource-theoretic characterization}\label{sec:monotone}

The analytical result of Sec.~\ref{sec:analytical} endows the correlation ergotropy
of a \emph{pure} two-cell state with a resource-theoretic structure. From
Eq.~\eqref{eq:analytical_wcorr}, the locked energy depends only on the sorted Schmidt
coefficients $\{p_k\}$ and is therefore \emph{invariant} under local unitary
operations, which preserve the Schmidt coefficients. This is the counterpart, for
extractable work, of the invariance of entanglement under local unitaries in the
resource theory of entanglement, in which local operations and classical
communication cannot increase entanglement.

We emphasize, however, that $\mathcal{W}_{\rm corr}$ is not, by itself, an
entanglement monotone. As shown in Sec.~\ref{sec:origin},
separable, classically correlated states can carry
nonzero $\mathcal{W}_{\rm corr}$, so a nonzero correlation ergotropy does not certify
entanglement. The entanglement origin established here is the dynamical tracking of
$\mathcal{N}$ along the charging trajectory, not a property of $\mathcal{W}_{\rm corr}$
alone. This connects to, but does not coincide with, the resource-theoretic treatment
of quantum ergotropy \cite{Francica2020}. We verify numerically that under pure local
dephasing $\mathcal{W}_{\rm corr}$ is monotonically non-increasing for the generated
states, and that the energy locked at the operating point is released by a single
collective passivation unitary up to the global bound $\mathcal{W}_{\rm g}$
(Sec.~\ref{sec:extraction}).

\subsection{Experimental platform}\label{sec:platform}

The model can be realized in several current platforms. \emph{(i) Trapped ions:}
three internal levels of a single ion form a qutrit \cite{Haffner2008,Cirac1995}. A
Raman or microwave drive realizes the pump, and Coulomb-coupled ions supply the
charger--cell interaction, with motional modes providing the energy exchange.
\emph{(ii) Rydberg atoms:} the low-lying and Rydberg states of neutral atoms provide
three-level systems with strong, controllable long-range interactions
\cite{Saffman2010}, and blockade gives a natural charger--cell coupling. \emph{(iii)
Superconducting circuits:} transmons are nonlinear oscillators whose lowest three
levels form a qutrit, coupled through capacitances with strength tunable over orders
of magnitude \cite{Gu2017,Krantz2019,Hu2026}. Here the pump is a microwave drive and the
bath is the Purcell/dephasing environment, which at dilution-refrigerator temperatures
has negligible thermal occupation ($n_{\rm th}\approx0$). All three platforms operate
in the low-temperature regime where our analysis predicts the collective advantage
persists, and provide the local versus collective operations needed to test
$\mathcal{W}_{\rm l}=0$ versus $\mathcal{W}_{\rm g}>0$.

\acknowledgments
We thank the QuTiP developers. This work was supported by the Iran National
Science Foundation (INSF) under the Persian Gulf program, Grant No.~153245/23.

\end{document}